# Reversible Photomechanical Switching of Individual Engineered Molecules at a Surface

Matthew J. Comstock[1,3], Niv Levy[1,3], Armen Kirakosian[1,3], Jongweon Cho[1,3], Frank Lauterwasser[2,3], Jessica H. Harvey[2,3], David A. Strubbe[1,3], Jean M.J. Fréchet[2,3], Dirk Trauner[2,3], Steven G. Louie[1,3], and Michael F. Crommie[1,3]

[1]Department of Physics, University of California at Berkeley, Berkeley, California 94720-7300

[2]Department of Chemistry, University of California, Berkeley, California 94720-1460

[3]Materials Sciences Division, Lawrence Berkeley National Laboratory, Berkeley, California 94720

We have observed reversible light-induced mechanical switching for individual organic molecules bound to a metal surface. Scanning tunneling microscopy (STM) was used to image the features of individual azobenzene molecules on Au(111) before and after reversibly cycling their mechanical structure between *trans* and *cis* states using light. Azobenzene molecules were engineered to increase their surface photomechanical activity by attaching varying numbers of tert-butyl (TB) ligands ("legs") to the azobenzene phenyl rings. STM images show that increasing the number of TB legs "lifts" the azobenzene molecules from the substrate, thereby increasing molecular photomechanical activity by decreasing molecule-surface coupling.

The conversion of light to mechanical motion at the molecular level provides exciting possibilities for nanomachine control and characterization, including high frequency and non-contact operation [1, 2]. Progress in this area has occurred through the investigation of solution-based molecular machine ensembles [3], organic polymers capable of light-induced expansion and contraction [4, 5], light-controlled ion channels [6], and other surface-molecule systems [7-9]. A common motif here is to employ molecular sub-units, such as azobenzene, known from ensemble measurements to reversibly transform from one isomeric state (e.g., *trans*) to another (e.g., *cis*) upon absorption of light (Fig. 1) [10]. A central concern is what strategies might be used to reversibly and optically control the mechanical state of a single, addressable molecule, and how such strategies are influenced by the coupling between a molecule and its environment.

In order to explore this issue, we have used scanning tunneling microscopy (STM) to spatially resolve the features of individual azobenzene molecules on a gold surface before and after reversibly cycling their mechanical structure between *cis* and *trans* states via photoisomerization. This procedure is different from previous STM tip-induced molecular manipulation studies [11-14] in that it is performed in the absence of an STM tip, it explores a different physical regime (i.e., photomechanical coupling), and it offers the flexible dynamical control inherent to optical processes.

We achieved reversible single-molecule photoisomerization by engineering azobenzene molecules to increase their surface photomechanical activity. While gas- and solution-phase azobenzene molecules readily photoisomerize [10], this process can be quenched at a surface by molecule-surface coupling [15-17]. We therefore attached *tert*- butyl "legs" (TB: $C_4H_9$) to an azobenzene scaffold ($C_{12}H_{10}N_2$) to reduce this coupling. When illuminated by UV light, azobenzene molecules made with zero or two TB legs did not photoisomerize when placed on a gold surface, but azobenzene molecules with four attached TB legs did. Single-molecule photoisomerization was confirmed unambiguously by the reversibility of the photoreaction and by comparing experimentally resolved intramolecular features of single *trans* and *cis* azobenzene molecules with *ab initio* simulations. The "transition" that we observe from quenched to active photomechanical behavior reveals the importance of electro-mechanical coupling between a molecule and substrate.

We performed our measurements using a home-built variable-temperature ultrahigh vacuum STM. Two-legged 4,4'-di-*tert*-butyl-azobenzene (DTB-azobenzene) and four-legged 3,3',5,5'-tetra-*tert*-butyl-azobenzene (TTB-azobenzene) were synthesized via oxidative coupling reactions of 4-*tert*-butyl-aniline and 3,5-di-*tert*-butyl-aniline, respectively [18]. *Trans* isomers of the molecules were deposited via leak valve and Knudsen cell techniques onto clean Au(111) substrates held at 30 K. Samples were then annealed at room temperature for 10 minutes in order to achieve ordered molecular arrangements. STM images were acquired in the temperature range of 25 K to 30 K using tunnel currents below 50 pA for stable imaging. A CW diode laser at an external viewport provided UV radiation at 375 nm with an average intensity of 90 mW/cm$^2$ at the sample surface. During UV exposures the STM tip was retracted and the sample temperature was maintained between 28 K and 32 K.

The STM images in Fig. 2 reveal the differences between adsorbed bare azobenzene (no TB legs), DTB-azobenzene (two TB legs), and TTB-azobenzene (four

TB legs). Bare azobenzene molecules (Fig. 2A) appear as pairs of closely touching lobes, with each lobe indicating the position of a single phenyl ring [11, 19]. Individual DTB-azobenzene molecules (Fig. 2B) similarly appear as a pair of lobes, except that the DTB-azobenzene lobes are separated by a wider gap. Individual TTB-azobenzene molecules (Fig. 2C) appear as four-lobed structures [14]. The appearance of DTB-azobenzene and TTB-azobenzene is consistent with their expected TB leg arrangements (see models in Figs. 2B-2C). Constant current linescans across the azobenzene derivatives (Fig. 2D) show that the DTB-azobenzene and TTB-azobenzene molecules are progressively taller than bare azobenzene molecules. Hence molecular engineering using TB-leg functionalization [20] achieves progressive "lifting" of photomechanical molecules away from a surface.

The photomechanical activity of this series of azobenzene derivatives was checked by illuminating each type of molecular adsorbate separately with an equal exposure to UV light. Successful UV-induced switching was observed only for the four-legged TTB-azobenzene molecules. Figure 3 shows the same island of TTB-azobenzene molecules on Au(111) before and after a three hour exposure to UV light. Before UV exposure the island is uniformly composed of the *trans* isomer. After UV exposure the emergence of new, bright protrusions can be seen in the island. While *trans*-TTB-azobenzene molecules display four peripheral lobes before UV illumination, the UV-transformed TTB-azobenzene molecules display only three peripheral lobes along with a new, bright (i.e., "tall") feature near the center of the molecule (Fig. 3 insets and Fig. 4). Approximately 4% of *trans*-TTB-azobenzene molecules photoswitch to the new "three-lobe state" after a 1 hour UV exposure at 90 mW/cm$^2$ [21].

The observed photomechanical switching can be optically reversed for single molecules by re-exposing the molecules to UV light. Fig. 4 shows one particular TTB-azobenzene molecule undergoing a complete cycle of reversible photo-switching (*trans* state → three-lobe state → *trans* state). The reversibility of the photoswitching provides strong evidence that the three-lobe state is indeed the *cis* isomer of TTB-azobenzene, since reversibility rules out other possible structural change mechanisms such as photodissociation (STM tip-based molecular manipulation in the absence of photons can also achieve similar structural transformations [14]).

*Ab initio* density functional theory (DFT) calculations predict TTB-azobenzene *cis* and *trans* isomer appearances very close to the experimentally observed molecules (Fig. 5). Local density of states (LDOS) calculations were performed for isolated *trans*- and *cis*-TTB-azobenzene molecules using the SIESTA code [22] (similarly to Ref. [23] but with the generalized-gradient approximation (GGA) [24]). The *trans* and *cis* isomer molecular structures were optimized via energy minimization (Fig. 5A and B, *cis* CNNC and CCNN angles are 11° and 47° respectively), and isosurfaces of HOMO orbital LDOS were calculated to simulate STM images. The simulated *trans* isomer STM image is dominated by four peripheral lobes at the TB leg positions (Fig. 5C). The simulated *cis* isomer STM image shows a bright central area due to the upwards rotation of one TB leg, leaving the three remaining TB legs on the periphery below (Fig. 5D). A simple tiling of the calculated *trans* and *cis* isomer simulated images using experimentally observed lattice positions (Fig. 5E) shows that the simulated *trans*- and *cis*-TTB-azobenzene images match the experimental data (Fig. 5F) quite well.

All of the azobenzene molecules studied here photoisomerize easily in solution [25] and so it is clear that the change in environment at a surface plays a role in modifying molecular photoswitching behavior. The influence of a surface on molecular photoswitching can be divided into three general mechanisms: (i) *Steric hindrance*: molecules lose the freedom to change conformation if atomic motion is constrained by either the surface or neighboring molecules [10]. (ii) *Electronic lifetime effects*: if the surface-modified lifetime of photo-excited electrons is less than the time it takes for a molecule to complete a conformational change then photoswitching can be quenched [10, 15, 17]. (iii) *Substrate-induced changes in optical absorption*: hybridization between an adsorbed molecule and a surface can change the optical absorption spectrum (and subsequent photoswitching properties) of a molecule [15]. A recent theoretical study proposes that, analogous to the Franck-Condon principle, coupling between molecular and substrate electromechanical degrees of freedom (e.g. dissipative modes such as phonons) can quench the optical absorption necessary for azobenzene *cis* ↔ *trans* photoswitching [17].

We believe that steric hindrance due to gold surface attachment is not the cause of quenching, as follows: azobenzene bonds weakly to gold (physisorption limit) [11], we observe similar surface diffusion rates for the different azobenzene derivatives, and STM tip-pulsing experiments show that it is possible to isomerize bare azobenzene on gold using tip-manipulation techniques [13]. Steric hindrance due to molecule-molecule interactions within islands is also not likely to play a dominant role in photoquenching. Molecule-molecule bonding appears weak because we can easily separate azobenzene molecules without damage using STM manipulation and we also measure a very low melting point (approximately 50 K) for islands of all three azobenzene species. Furthermore, molecules at the boundaries of islands (which have fewer nearest neighbors compared to interior molecules) show identical photoswitching to interior molecules.

Our observation of an increase in azobenzene photoswitching rate as the molecule is "lifted" off the surface is consistent with quenching mechanisms (ii) and (iii). We find these possibilities difficult to distinguish at present. Molecule-substrate electronic state hybridization can lead to both optical absorption shifts and changes in excited state lifetimes [15]. Recent theoretical work, however, predicts that increased hybridization of azobenzene to substrate dissipative modes will cause a sharp transition to completely quenched photoisomerization accompanied by gaps opening in optical absorption bands [17]. Future STM measurements of the dependence of single-molecule photoswitching on the wavelength of light, as well as new optical absorption measurements, will be useful in distinguishing the mechanisms that dominate photoswitching in this hybrid molecule-surface system.

In conclusion, we have experimentally observed reversible photomechanical switching for individual azobenzene molecules at a metal surface. Our measurements reveal the significance of environmental coupling in determining molecular photo-switching behavior. This effect will likely play an important role in future applications of molecular photoswitching in nanostructured condensed matter systems.

ACKNOWLEDGEMENT
We thank Carine Edder for her assistance in early experiments and K. H. Khoo for her assistance with simulations. This work was supported by the Director, Office of Science,

Figure Captions:

Fig. 1. Azobenzene photoisomerization reaction.

Fig. 2. STM constant-current images of functionalized azobenzene molecules on Au(111) (T = 30 K, V = -1 V, I = 25 pA, images are scaled identically): (A) bare azobenzene, (B) DTB-azobenzene, (C) TTB-azobenzene. Upper panels show chemical structure of the *trans* isomers of the imaged molecules. Single molecule images are identified by white boxes in insets (dotted line in each box shows linescan trajectory for (D)). (D) Linescans across different functionalized molecules show apparent height on Au(111). DTB-azobenzene and TTB-azobenzene linescans were taken at the edge of islands. Dashed part of linescan provides guide-to-the-eye for identifying single-molecule width.

Fig. 3. Photoisomerization of individual TTB-azobenzene molecules on Au(111) from *trans* to *cis*. Same island of TTB-azobenzene molecules is shown before (upper image) and after (lower image) a three hour exposure to 90 mW/cm$^2$ UV irradiation at 375 nm. After UV exposure 45 TTB-azobenzene molecules have switched from the *trans* to the *cis* state. Inset zoom-in images show UV-induced switching (before and after) from *trans* to *cis* for a single molecule (identified by white box).

Fig. 4. Reversible photo-induced switching is observed for a single TTB-azobenzene molecule. The same individual TTB-azobenzene molecule (identified by white boxes in three successive panels) is shown before and after two successive exposures to UV light. The molecule starts out in the *trans* state (top panel), is then switched to the *cis* state after the first exposure to UV light (middle panel), and is then switched back to the *trans* state after a second exposure to UV light (bottom panel).

Fig. 5. Simulated *trans*- and *cis*-TTB-azobenzene structures compared to experiment. (A) Calculated *trans* geometry. (B) Calculated *cis* geometry. (C) Calculated *trans* local density of states (LDOS) integrated from $E_F$ to $E_F - 1$ eV, at an isosurface about 3 Å away from the nearest atoms. (D) Calculated *cis* LDOS isosurface (same parameters as in (C)). (E) Simulated STM image of TTB-azobenzene using tiled single-molecule LDOS isosurfaces from (C) and (D) (image has been smoothed using a 0.2 nm width Gaussian blur filter to approximate experimental convolution with the STM tip). (F) Experimental STM image of TTB-azobenzene molecules including one photoisomerized *cis* isomer.

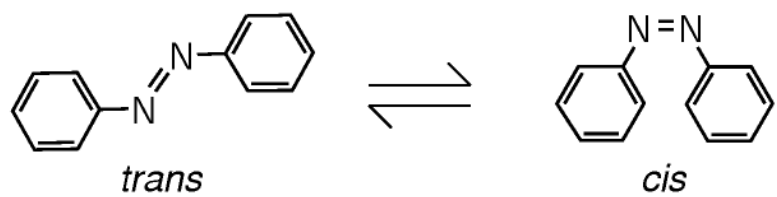

Figure 1

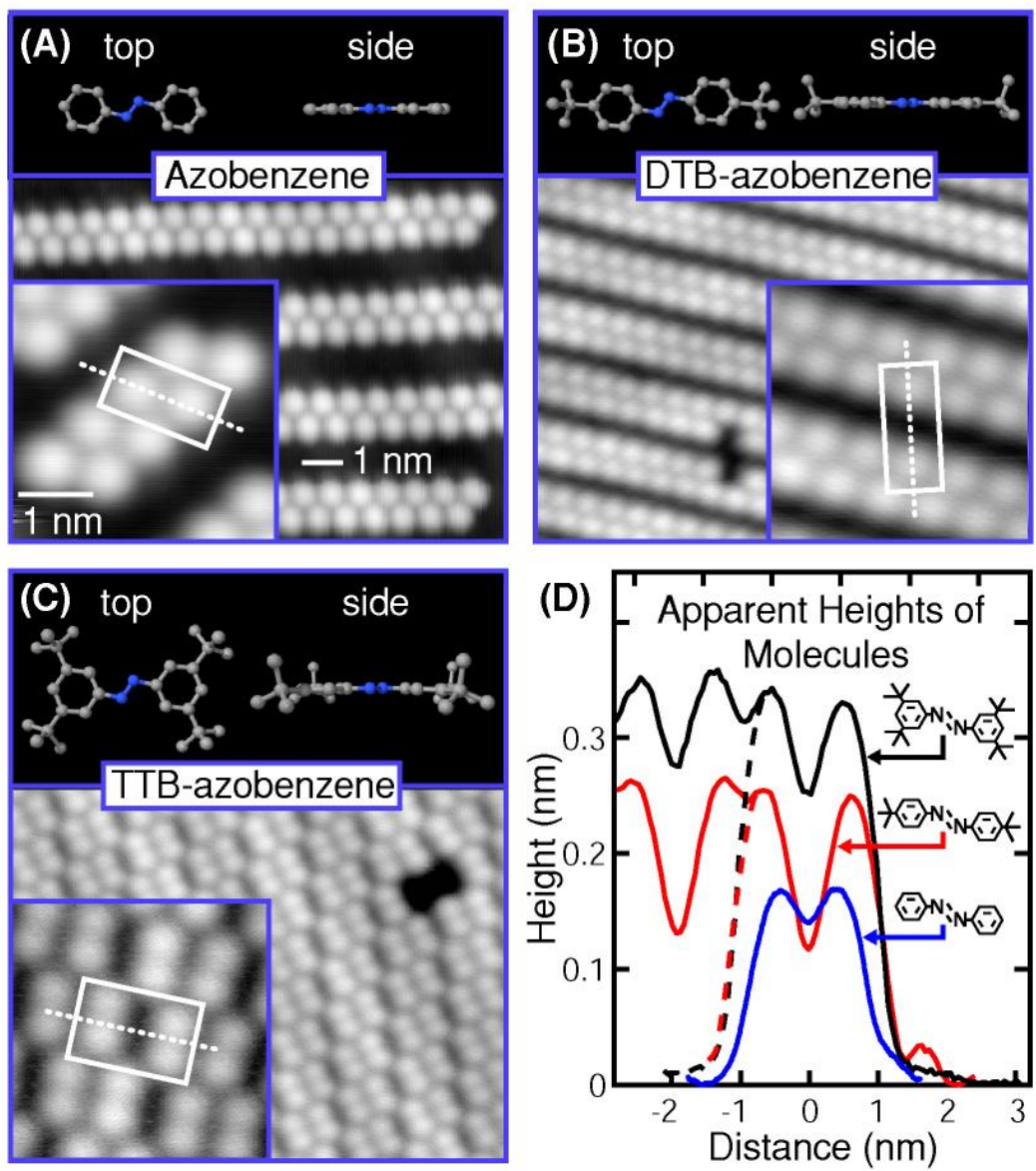

Figure 2

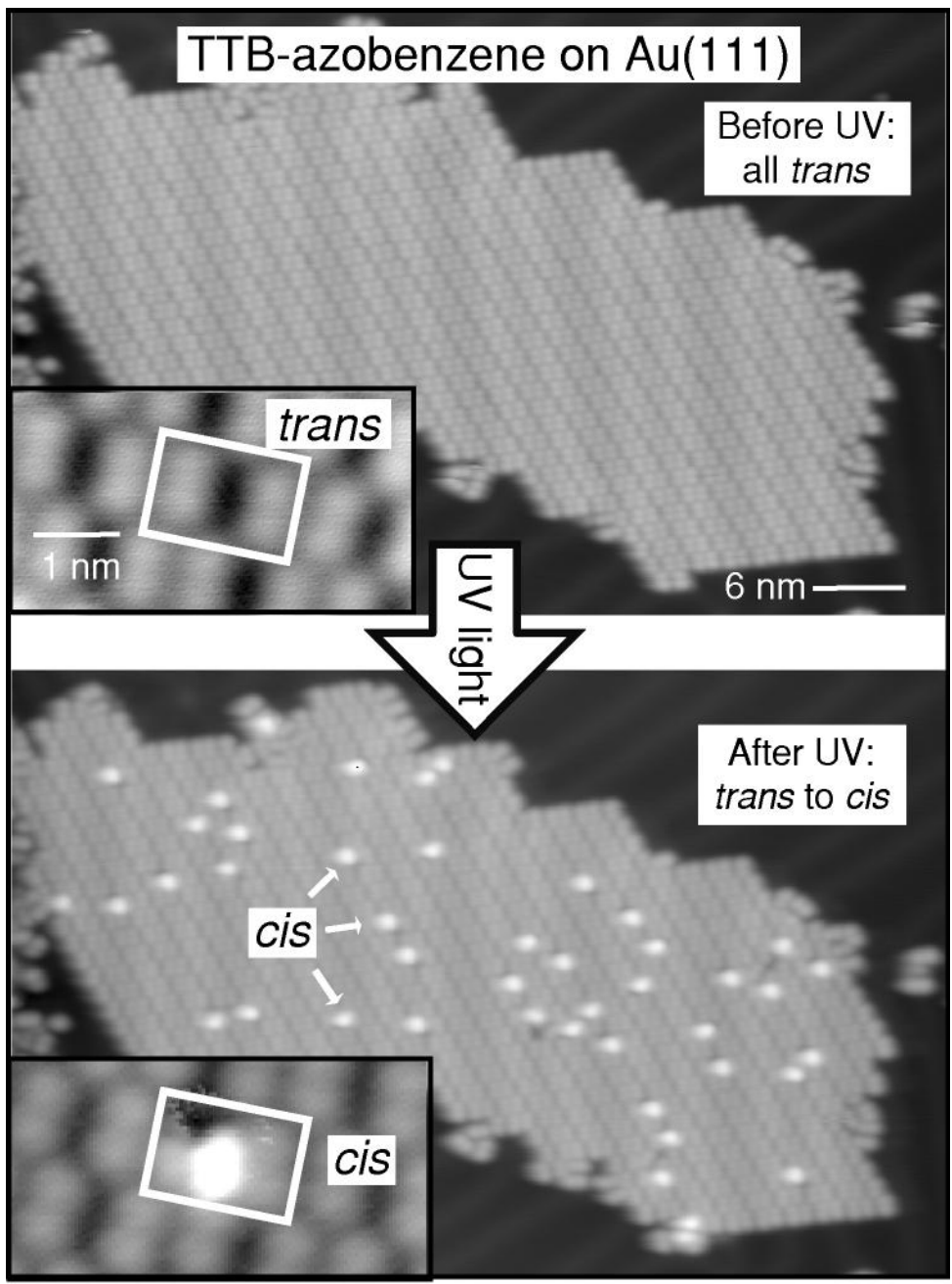

Figure 3

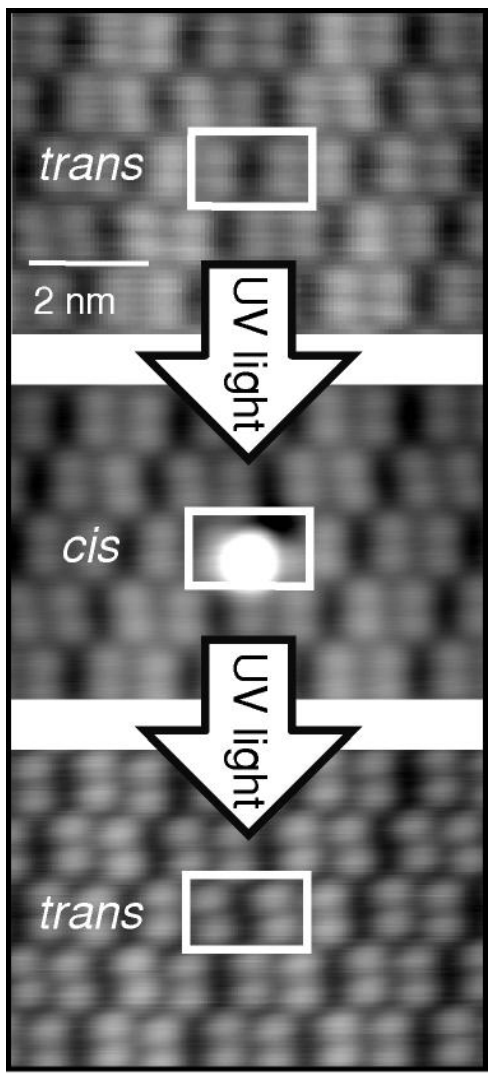

Figure 4

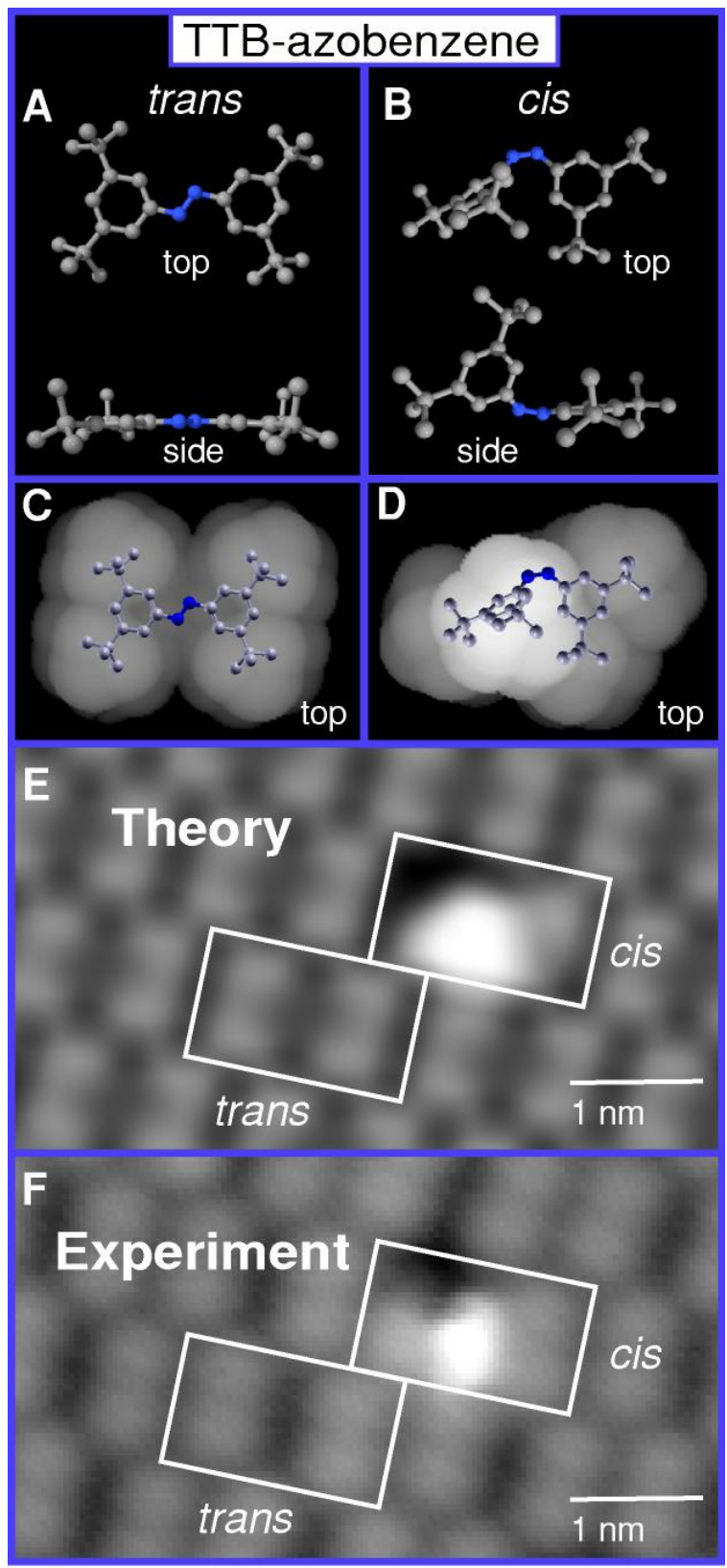

Figure 5